\documentclass[conference]{IEEEtran}
\usepackage{cite}

\ifCLASSINFOpdf
\usepackage[pdftex]{graphicx}
\else
\fi
\usepackage{amsmath}
\usepackage{amsfonts} 
\usepackage{amsthm}
\usepackage{amssymb}
\usepackage[boxruled]{algorithm2e}
\usepackage{amsthm} 
\usepackage{mathtools} 
\usepackage{textcomp} 

\usepackage{BOONDOX-ds}

\usepackage{bbm}

\newcommand{\pfun}{\mathop{\hbox{$\to$\kern-7pt\raise.9pt\hbox{\scalebox{1}[.55]{$|$}}\kern4pt} }}

\usepackage{array}

\begin{document}

\title{Vertical, Temporal, and Horizontal Scaling of Hierarchical Hypersparse GraphBLAS Matrices}

\author{\IEEEauthorblockN{Jeremy Kepner$^{1,2,3}$, Tim Davis$^{4}$, Chansup Byun$^1$, William Arcand$^1$, David  Bestor$^1$, William Bergeron$^1$, \\
Vijay Gadepally$^{1,2}$, Michael Houle$^1$, Matthew Hubbell$^1$, Michael Jones$^1$, Anna Klein$^1$, Lauren
    Milechin$^5$, \\ Julie Mullen$^1$, Andrew Prout$^1$, Albert Reuther$^1$, Antonio Rosa$^1$,   Siddharth Samsi$^1$, Charles Yee$^1$, Peter Michaleas$^1$
\\
\IEEEauthorblockA{$^1$MIT Lincoln Laboratory Supercomputing Center, $^2$MIT Computer Science \& AI Laboratory, \\ $^3$MIT Mathematics Department, $^4$Texas A\&M, $^5$MIT Department of Earth, Atmospheric and Planetary Sciences}}}
\maketitle

\begin{abstract}
Hypersparse matrices are a powerful enabler for a variety of network, health, finance, and social applications.  Hierarchical hypersparse GraphBLAS matrices enable rapid streaming updates while preserving algebraic analytic power and convenience.   In many contexts, the rate of these updates sets the bounds on performance.  This paper explores hierarchical hypersparse update performance on a variety of hardware with identical software configurations.  The high-level language bindings of the GraphBLAS readily enable performance experiments on simultaneous diverse hardware.  The best single process performance measured was 4,000,000 updates per second.  The best single node performance measured was 170,000,000 updates per second. The hardware used spans nearly a decade and allows a direct comparison of hardware improvements for this computation over this time range; showing a 2x increase in single-core performance, a 3x increase in single process performance, and a 5x increase in single node performance.   Running on nearly 2,000  MIT SuperCloud nodes simultaneously achieved a sustained update rate of over 200,000,000,000 updates per second.  Hierarchical hypersparse GraphBLAS allows the MIT SuperCloud to analyze extremely large streaming network data sets.  
\end{abstract}


%
\IEEEpeerreviewmaketitle

\section{Introduction}
\let\thefootnote\relax\footnotetext{This material is based upon work supported by the Assistant Secretary of Defense for Research and Engineering under Air Force Contract No. FA8702-15-D-0001, National Science Foundation CCF-1533644, and United States Air Force Research Laboratory Cooperative Agreement Number FA8750-19-2-1000. Any opinions, findings, conclusions or recommendations expressed in this material are those of the author(s) and do not necessarily reflect the views of the Assistant Secretary of Defense for Research and Engineering, the National Science Foundation, or the United States Air Force. The U.S. Government is authorized to reproduce and distribute reprints for Government purposes notwithstanding any copyright notation herein.}

Streaming data plays a critical role in protecting computer networks, tracking the spread of diseases, optimizing financial markets, and ad-placement in social media.  The volumes of data in these applications is significant and growing.  For example, the global Internet is expected to exceed 100 terabytes per second (TB/s) creating significant performance challenges for the monitoring necessary to improve, maintain, and protect the Internet \cite{cisco2018cisco, allcott2017social, BadBotReport, ClaffyClark2020, kepner2021zero}.  As in other applications hypersparse matrices are a natural way to represent network traffic and rapidly constructing these traffic matrix databases is a significant productivity, scalability, representation, and performance challenge \cite{castellana2017high, busato2018hornet, 8547563, 8547570, 8547514, 8547572, samsi2017static, kao2017streaming, kepner2020multi}.

The SuiteSparse GraphBLAS library is an OpenMP accelerated C implementation of the GraphBLAS.org sparse matrix standard \cite{7761646, Davis:2019:ASG:3375544.3322125, 8547538} and makes widely available the benefits of hypersparse capabilities \cite{buluc2008representation, kepner2011graph, bulucc2012parallel}.  Python, Julia, and Matlab/Octave bindings allow the performance benefits of the SuiteSparse GraphBLAS C library to be realized in these highly productive programming environments.  Our team has developed a high-productivity scalable platform---the MIT SuperCloud---for providing scientists and engineers the tools they need to analyze large-scale dynamic data \cite{kepner2012dynamic, gadepally2018hyperscaling, 8547629}.  The MIT SuperCloud provides interactive analysis capabilities  accessible from high level programming environments (Python, Julia, Matlab/Octave) that scale to thousands of processing nodes.  MIT SuperCloud maintains a diverse set of hardware running an identical software stack that allows direct comparison of hardware from different eras.

This paper builds on our prior work \cite{kepner202075} with hierarchical hypersparse GraphBLAS matrices by more thoroughly exploring the vertical, temporal, and horizontal scaling performance of hierarchical hypersparse construction.  The vertical benchmarking explores how different numbers of GraphBLAS processes and threads perform on different multicore compute nodes.  The temporal benchmarking compares the performance of compute nodes from different eras.  The horizontal benchmarking examines the aggregate performance achieved by running on thousands of diverse compute nodes.

\begin{figure*}[]
\centering
\includegraphics[width=5.5in]{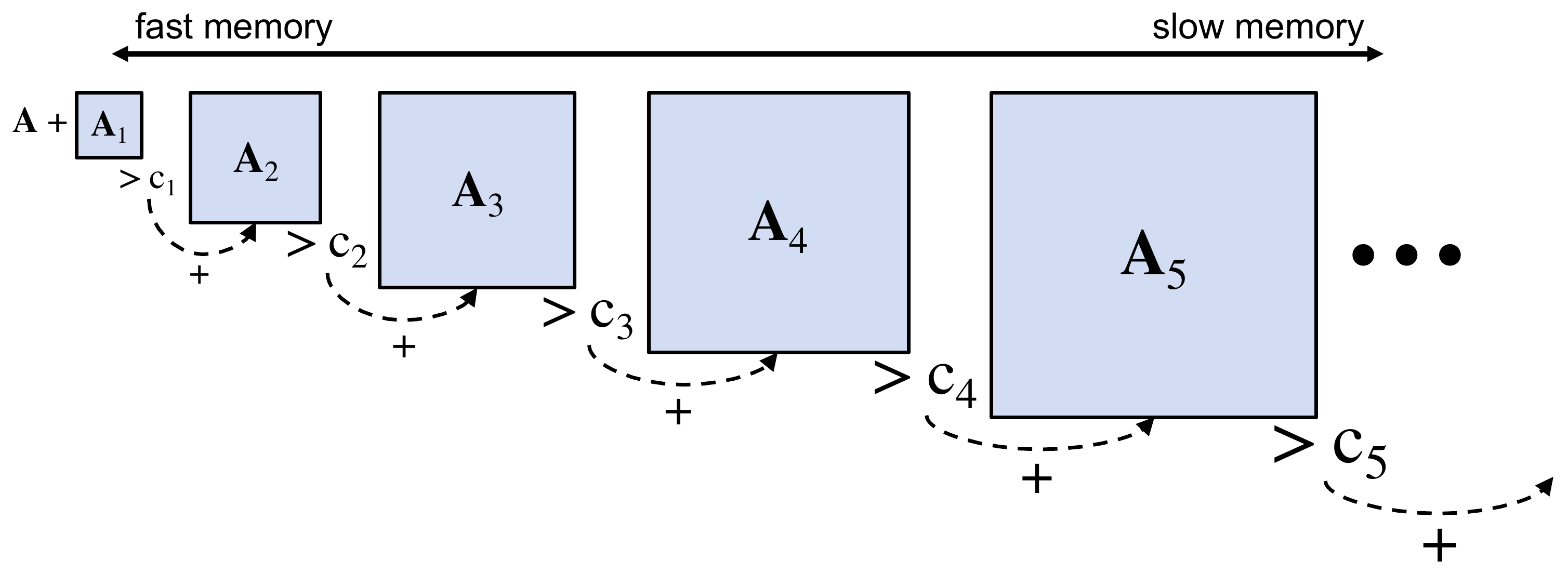}
\caption{Hierarchical hypersparse matrices store increasing numbers of nonzero entries in each layer (adapted from \cite{8916508}).  If the number of nonzero (stored) entries in layer ${\bf A}_i$ surpasses the nonzero threshold count $c_i$ then ${\bf A}_i$ is added to ${\bf A}_{i+1}$ and ${\bf A}_i$ is cleared.  Hierarchical hypersparse matrices ensure that the majority of updates are performed in fast memory.}
\label{fig:HierarchicalArrays}
\end{figure*}

\section{Hierarchical Hypersparse Matrices}

  Streaming updates to a large hypersparse matrix can be accelerated with a hierarchical implementation optimized to the memory hierarchy (see Fig.~\ref{fig:HierarchicalArrays}).   Rapid updates are performed on the smallest hypersparse matrices in the fastest memory.  The strong mathematical properties of the GraphBLAS allow a hierarchical implementation of hypersparse matrices to be implemented via simple addition. All creation and organization of hypersparse row and column indices are handled naturally by the GraphBLAS mathematics.  Hierarchical matrices are implemented with a simple heuristic. If the number of nonzero (nnz) entries in the matrix ${\bf A}_i$ at layer $i$ exceeds the cut threshold $c_i$, then ${\bf A}_i$ is added to ${\bf A}_{i+1}$ and ${\bf A}_i$ is cleared.  The overall usage is as follows
\begin{itemize}
\item Initialize $N$-level hierarchical hypersparse matrix with cuts $c_i$
\item Update by adding new data ${\bf A}$ to lowest layer
$$
  {\bf A}_1 = {\bf A}_1 + {\bf A}
$$
\item If ${\rm nnz}({\bf A}_1) > c_1$, then
$$
  {\bf A}_2 = {\bf A}_2 + {\bf A}_1
$$
and reset ${\bf A}_1$ to an empty hypersparse matrix of appropriate dimensions.
\end{itemize}

\noindent The above steps are repeated until ${\rm nnz}({\bf A}_i) \leq c_i$ or $i = N$.  To complete all pending updates for analysis, all the layers are added together
$$
   {\bf A}_{\rm all} = \sum_{i=1}^{N} {\bf A}_i
$$
Hierarchical hypersparse matrices dramatically reduce the number of updates to slow memory.  Upon query, all layers in the hierarchy are summed into the hypersparse matrix.  The cut values $c_i$ can be selected so as to optimize the performance with respect to particular applications.  The majority of the  updating is performed by using the existing GraphBLAS addition operation.  The corresponding Matlab/Octave GraphBLAS code for performing the update is a direct translation of the above mathematics as follows

\noindent \rule{\columnwidth}{0.5pt}
{\tt
\noindent function Ai = HierAdd(Ai,A,c);

   Ai\{1\} = Ai\{1\} + A;
   
   for i=1:length(c)
   
   ~  if (GrB.entries(Ai\{i\}) > c(i))
     
   ~~~    Ai\{i+1\} = Ai\{i+1\} + Ai\{i\};
       
   ~~~    Ai\{i\} = Ai\{length(c)+2\};
       
   ~  end
     
   end

\noindent end 
}

\noindent \rule{\columnwidth}{0.5pt}

\noindent The goal of hierarchical arrays are to manage the memory footprint of each level, so
the GraphBLAS {\tt GrB.entries()} command returns the number of entries in the GraphBLAS hypersparse matrix, which may include some materialized zero values.  In addition, {\tt GrB.entries()} executes much faster than the GraphBLAS number of nonzeros command {\tt nnz()}.  The last  hypersparse matrix stored in the hierarchical array is empty and is used to reinitialize layers whose entries have been cascaded to a subsequent layer.

\section{Computer Hardware}

\begin{table}
\caption{Computer Hardware Specifications}
\vspace{-0.25cm}
MIT SuperCloud maintains a diverse set of hardware running an identical modern software stack providing an unique platform for comparing performance over different eras.
\begin{center}
\includegraphics[width=\columnwidth]{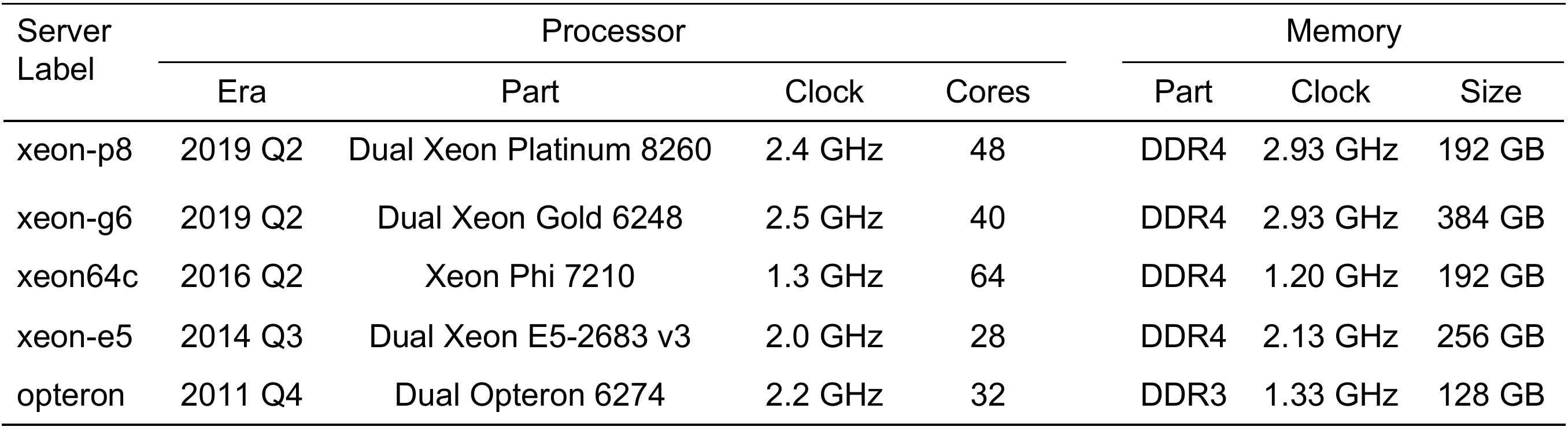}
\end{center}
\label{tab:HardwareTable}
\end{table}%

The GraphBLAS benchmarking  code was implemented using the Matlab/Octave bindings with the pMatlab parallel library \cite{Kepner2009}.  A typical run could be launched in a few seconds using the MIT SuperCloud triples-mode hierarchical launching system \cite{8547629}.  The launch parameters were [Nnodes Nprocess Nthread], corresponding to Nnodes nodes, Nprocess Matlab/Octave processes per node, and Nthread OpenMP threads per process.  On each node, each of the Nprocess processes and their corresponding Nthread threads were pinned to adjacent cores to minimize interprocess contention and maximize cache locality for the GraphBLAS OpenMP threads \cite{byun2019optimizing}.  Within each Matlab/Octave process, the underlying GraphBLAS OpenMP parallelism is used.  At the end of the processing the results were aggregated using asynchronous file-based messaging \cite{byun2019large}. Triples mode makes it easy to explore horizontal scaling  across nodes, vertical scaling by examining combinations of processes and threads on a node, and temporal scaling by running on diverse hardware from different eras.

The computing hardware consists of five different types of nodes acquired over a decade (see Table~\ref{tab:HardwareTable}).  The nodes are all multicore x86 compatible with comparable total memory.  The MIT SuperCloud maintains the same modern software across all nodes, which allows for direct comparison of hardware performance differences.

\section{Parameter Tuning}

\begin{figure}[]
\centering
\includegraphics[width=\columnwidth]{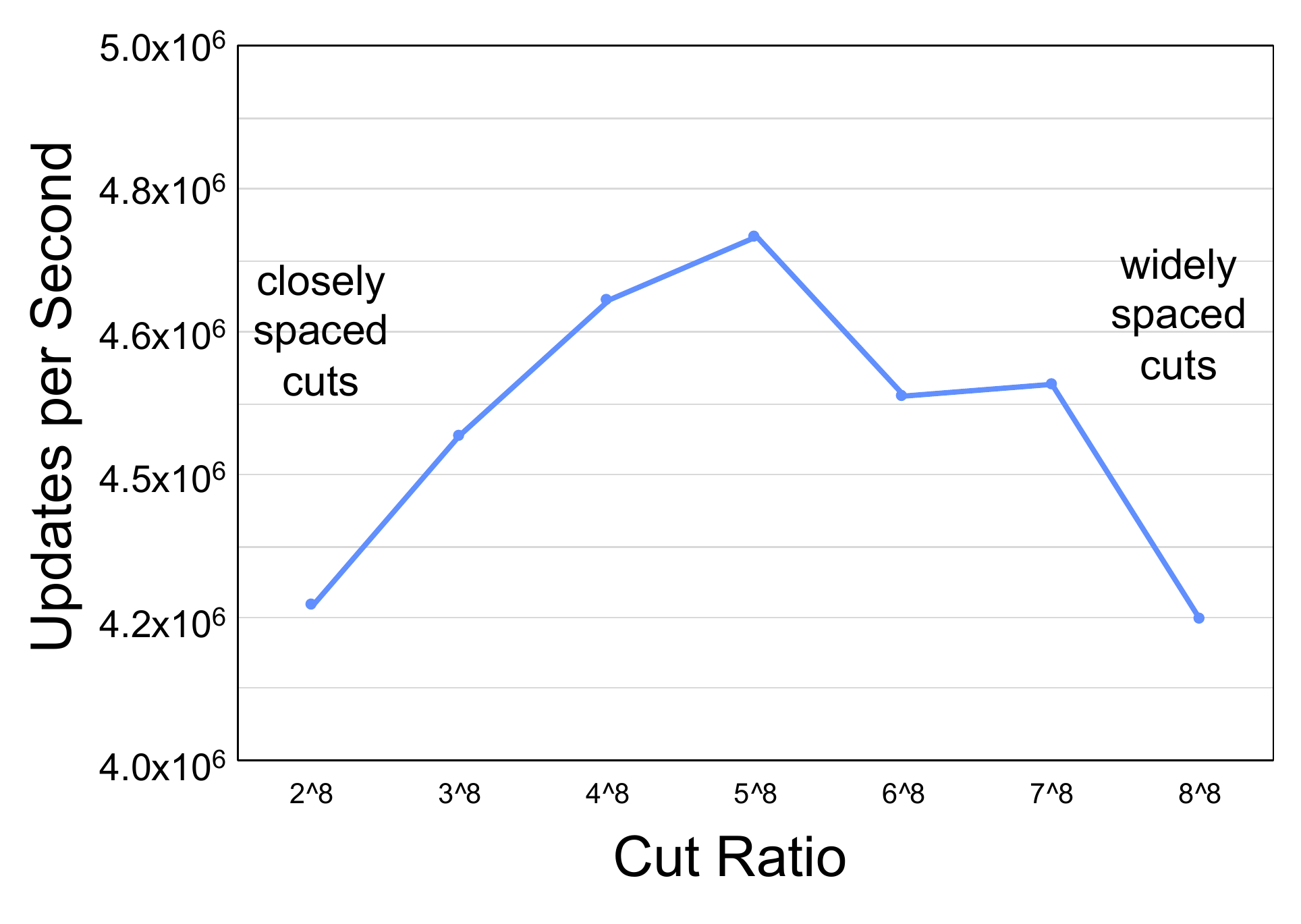}
\includegraphics[width=\columnwidth]{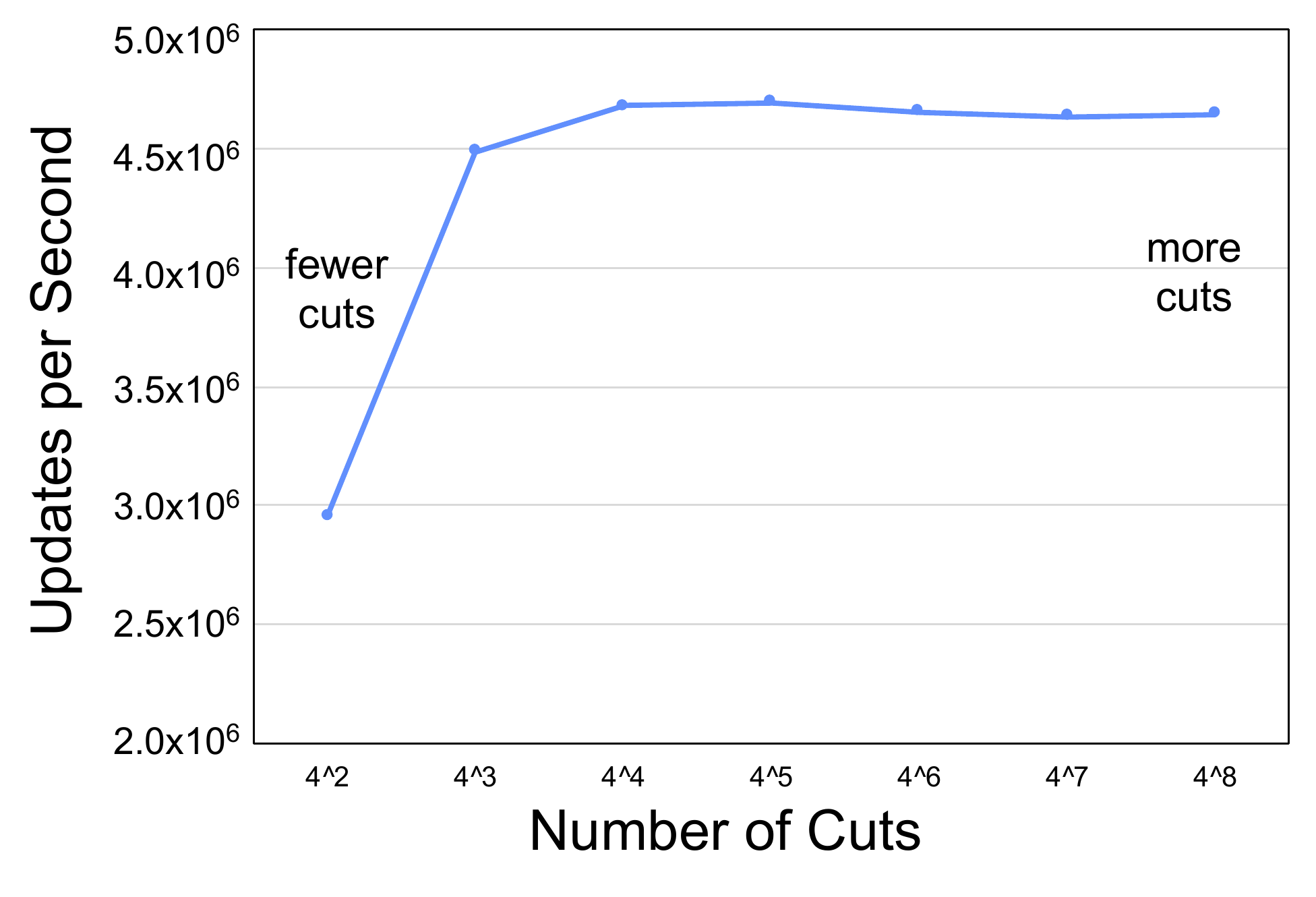}
\caption{Single thread update rate on a 48-core Dual Intel Xeon Platinum 8260 server for different cut ratio sets (top), ranging from $\{2^2, \ldots, 2^8\}$  to $\{8^2, \ldots, 8^8\}$, and different numbers of cuts (bottom).  For these data, performance is optimal with ratio spacings in the $3$ to $6$ range.  Actual cut values are determine by multiplying the cut ratios by a base value (in this case $2^{17}$).
}
\label{fig:CutValue}
\end{figure}

The performance of a hierarchical GraphBLAS for any particular problem is determined by the number of layers $N$ and the cut values $c_i$.  The parameters are tuned to achieve optimal performance for a given problem. These parameters were explored using simulated Graph500.org R-Mat power-law network data containing 100,000,000 connections that are inserted in groups of 100,000. Examples of different sets of cut values with different $c_1$  and different ratios between cut values are shown in Figure~\ref{fig:CutValue}.  These sets of cut values allow exploration of the update performance of  closely spaced cuts versus  widely spaced cuts.  For both $N$ and $c_i$ there are fairly broad ranges of comparable optimal performance.

\section{Vertical Scaling}

\begin{figure*}[]
\centering
\includegraphics[width=\columnwidth]{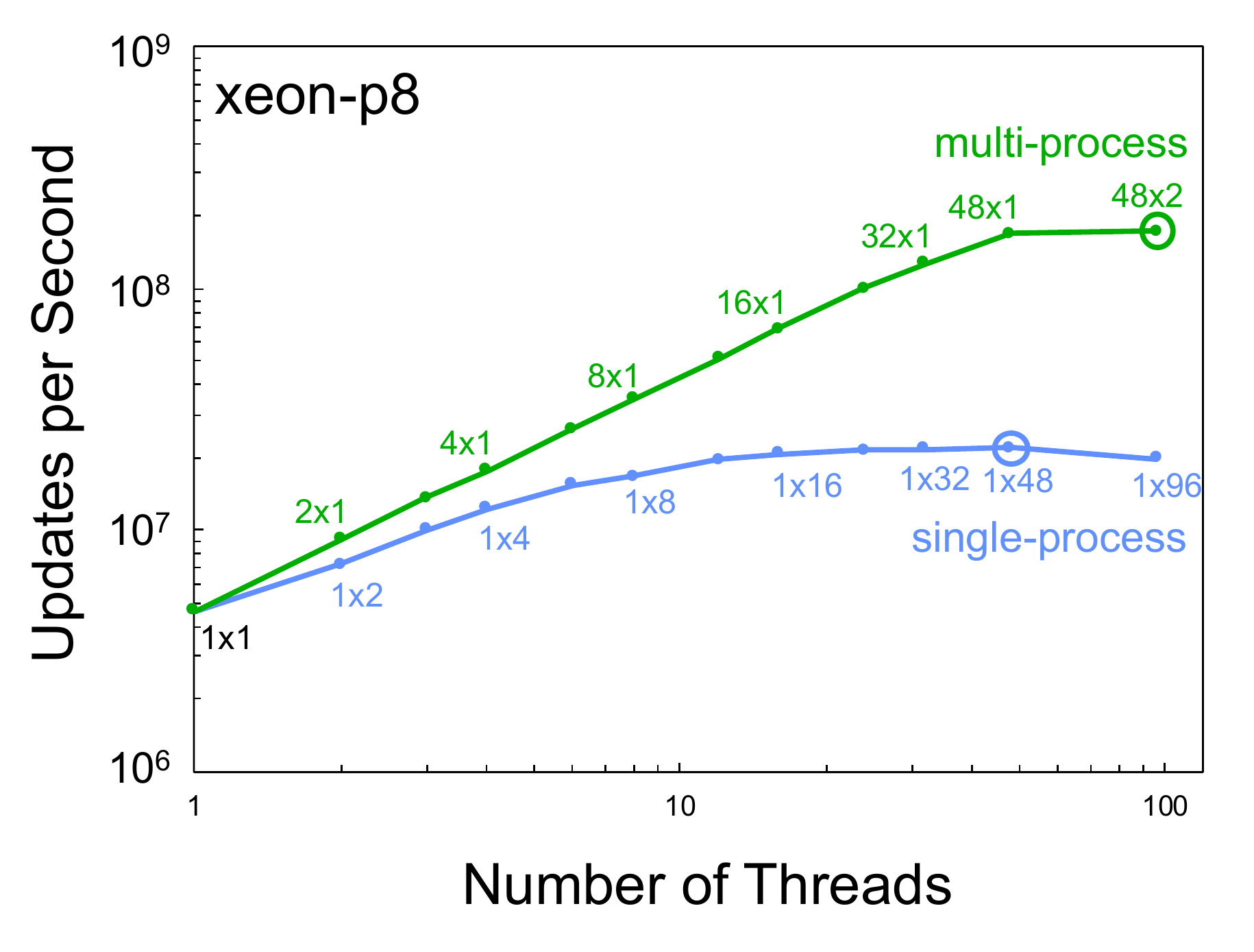}
\includegraphics[width=\columnwidth]{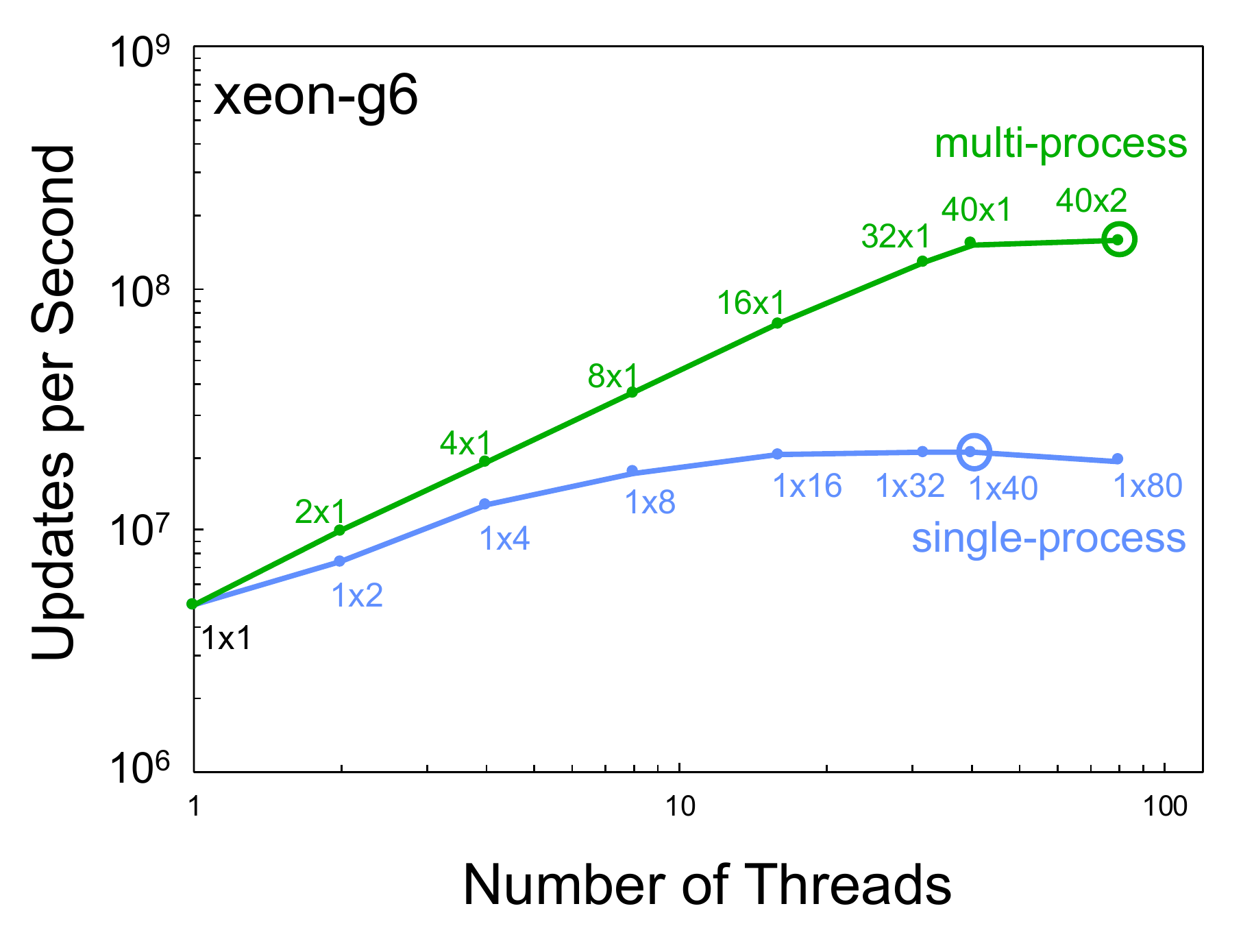}
\includegraphics[width=\columnwidth]{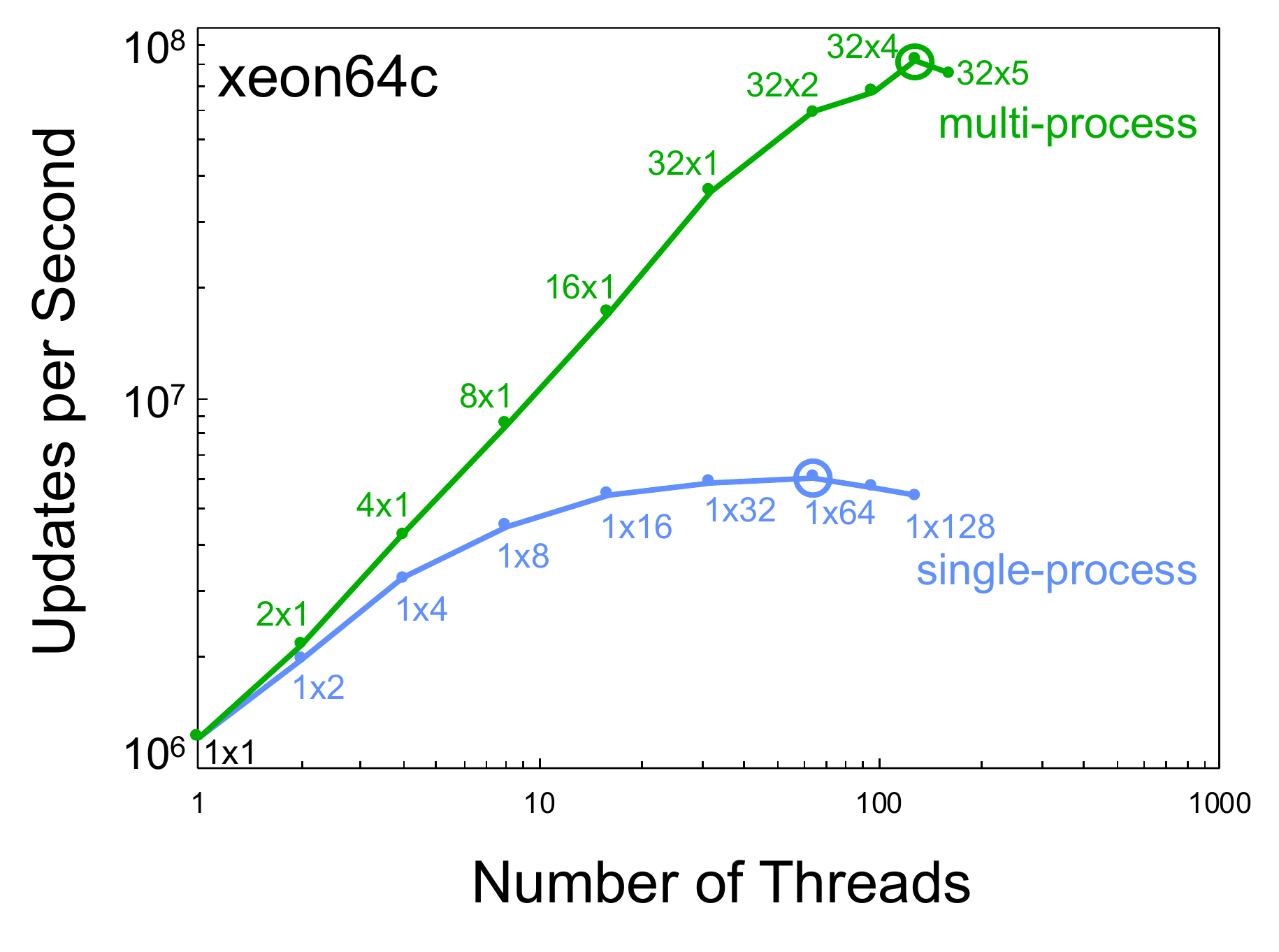}
\includegraphics[width=\columnwidth]{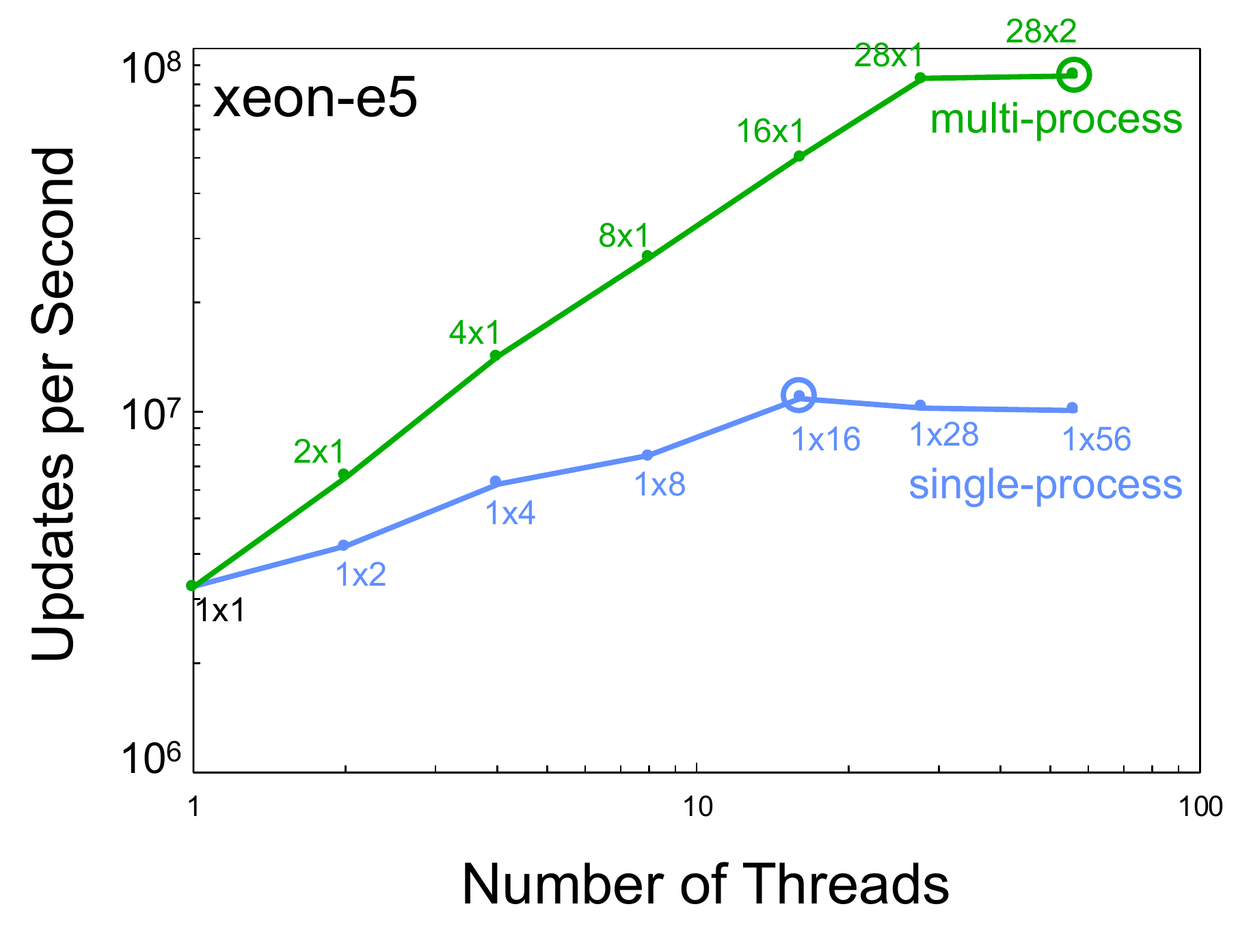}
\includegraphics[width=\columnwidth]{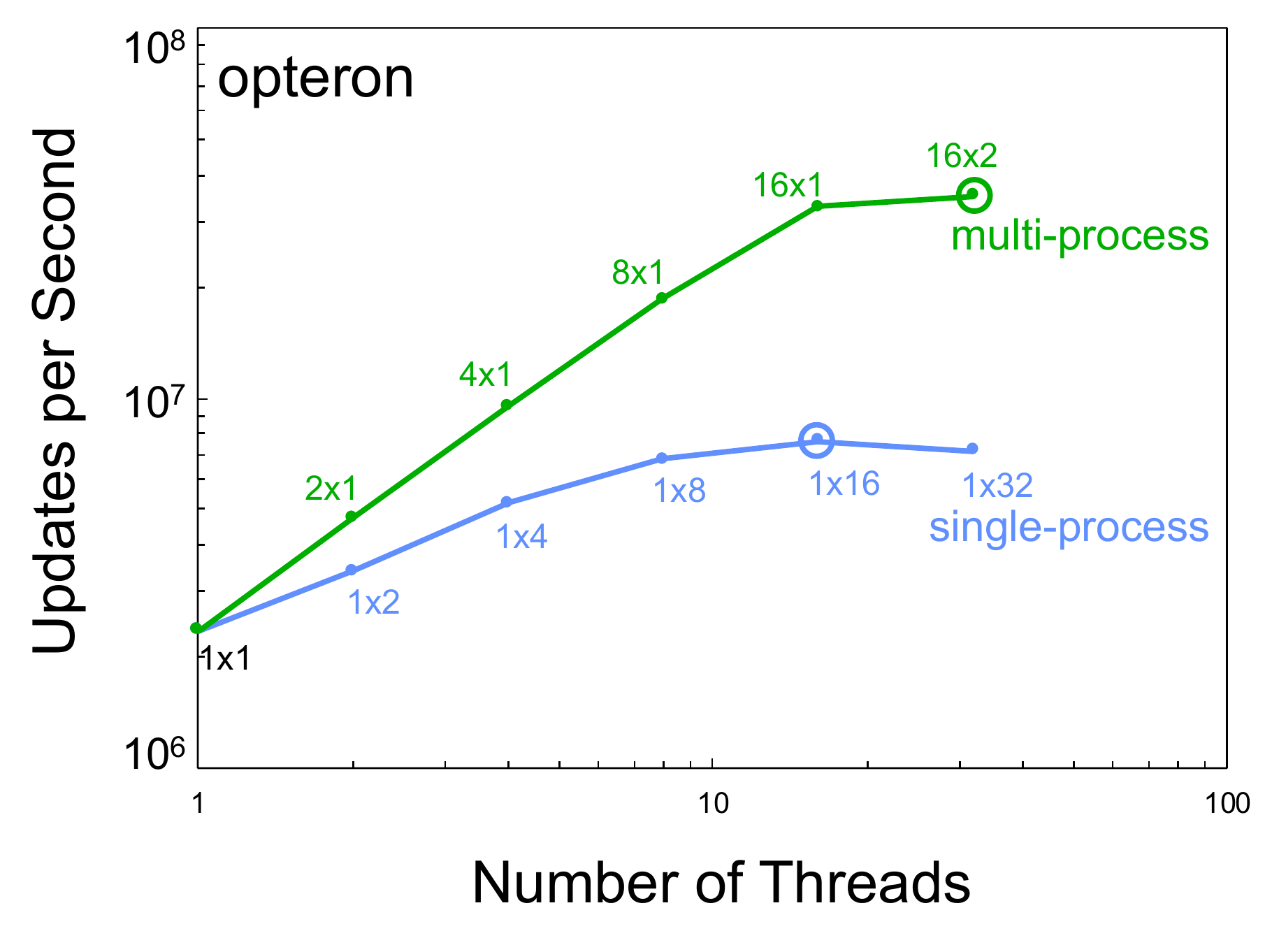}
\caption{Single node performance with different configurations of hierarchical GraphBLAS processes and threads on a xeon-p8 (upper left), xeon-g6 (upper right), xeon64c (middle left), xeon-e5 (middle right), and opteron (bottom).  Each data point is labeled: (\#processes)x(\#threads/process). The maximum performance for each case is denoted by a circle.}
\label{fig:UpdateRateSingleNode}
\end{figure*}

Vertical benchmarking explores how different numbers of GraphBLAS processes and threads perform on different multicore compute nodes.  This type of benchmarking is generally useful in most projects as it allows the determination of the best combination processing and threads prior to significant computation.  The number of processes and threads used for a given benchmark is denoted by Nprocess $\times$ Nthreads whose product is equal to total number threads used in the computation.  The GraphBLAS computation described in the previous section was repeated for two sets of parameters: single-process and multi-process.  In the single-process case the parameters tested are
$$
  1{\times}1, 1{\times}2, 1{\times}4, 1{\times}8, ...  
$$
In the multi-process case the parameters tested are
$$
  1{\times}1, 2{\times}1, 4{\times}1, 8{\times}1, ...  
$$
Figure~\ref{fig:UpdateRateSingleNode} shows the single node performance using different numbers of processes and threads for the different servers listed in Table~\ref{tab:HardwareTable}.  In all cases, the multi-process scaling provided greater aggregate performance and the single-process scaling provided a maximum of a 4x speedup over $1{\times}1$ case.

\section{Temporal Scaling}

\begin{figure}[]
\centering
\includegraphics[width=\columnwidth]{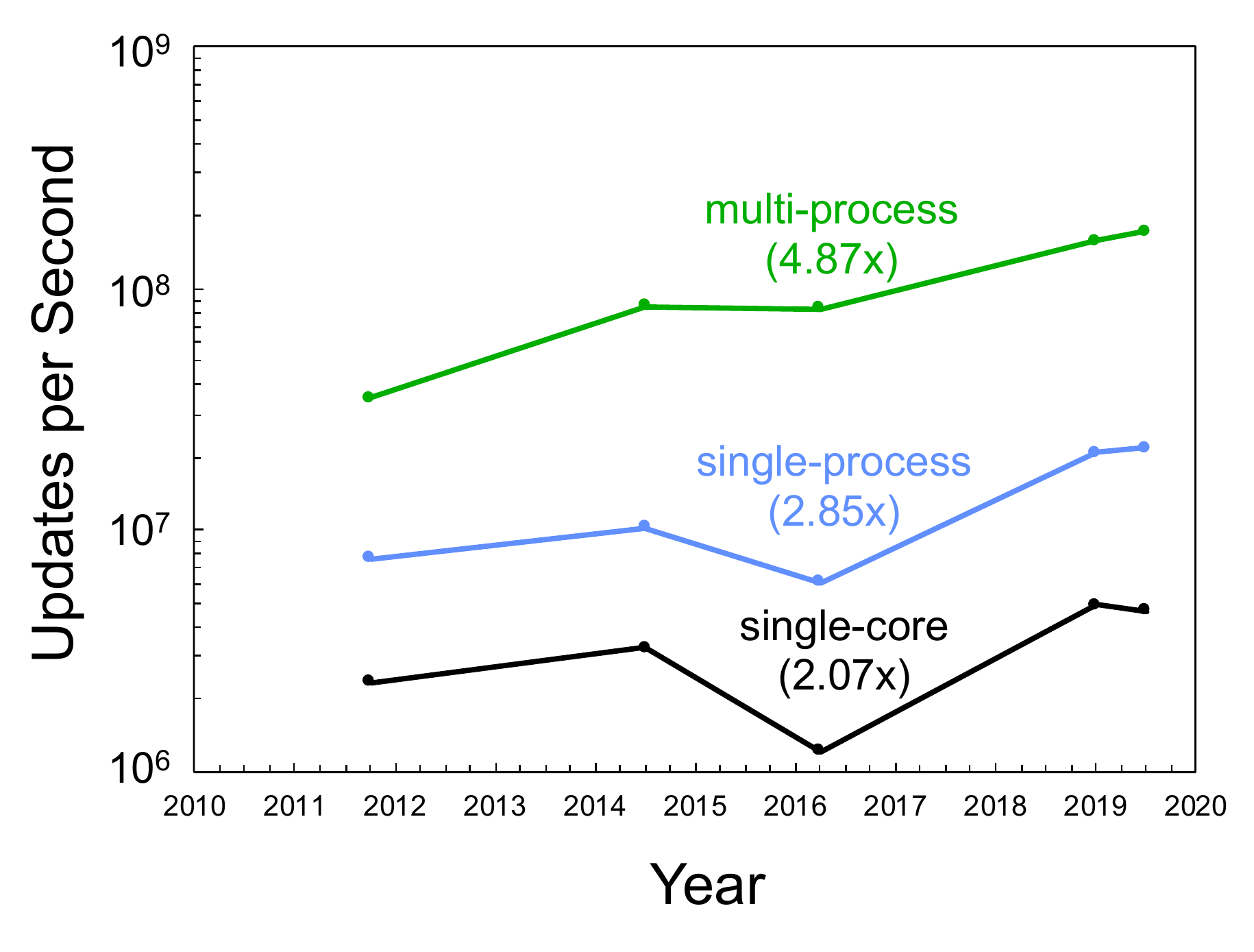}
\caption{Performance from hardware at different eras for a single process on a single core running a single thread (bottom black line), single process on a multiple cores running multiple threads (middle blue line), multiple process on a multiple cores each running a single threads (top green line).}
\label{fig:SingleNode-Year}
\end{figure}

Temporal benchmarking compares the performance of computing hardware from different eras.  The MIT SuperCloud provides a unique ability to directly compare hardware from different eras using exactly the same modern software stack.  Figure~\ref{fig:SingleNode-Year} shows the single core, best single-process, and best multi-process performance taken from Figure~\ref{fig:UpdateRateSingleNode} and plotted versus the year the computing hardware became available. These data show a 2x increase in single-core performance, a 3x increase in single process performance, and a 5x increase in single node performance.

\section{Horizontal Scaling}

\begin{figure}[]
\centering
\includegraphics[width=\columnwidth]{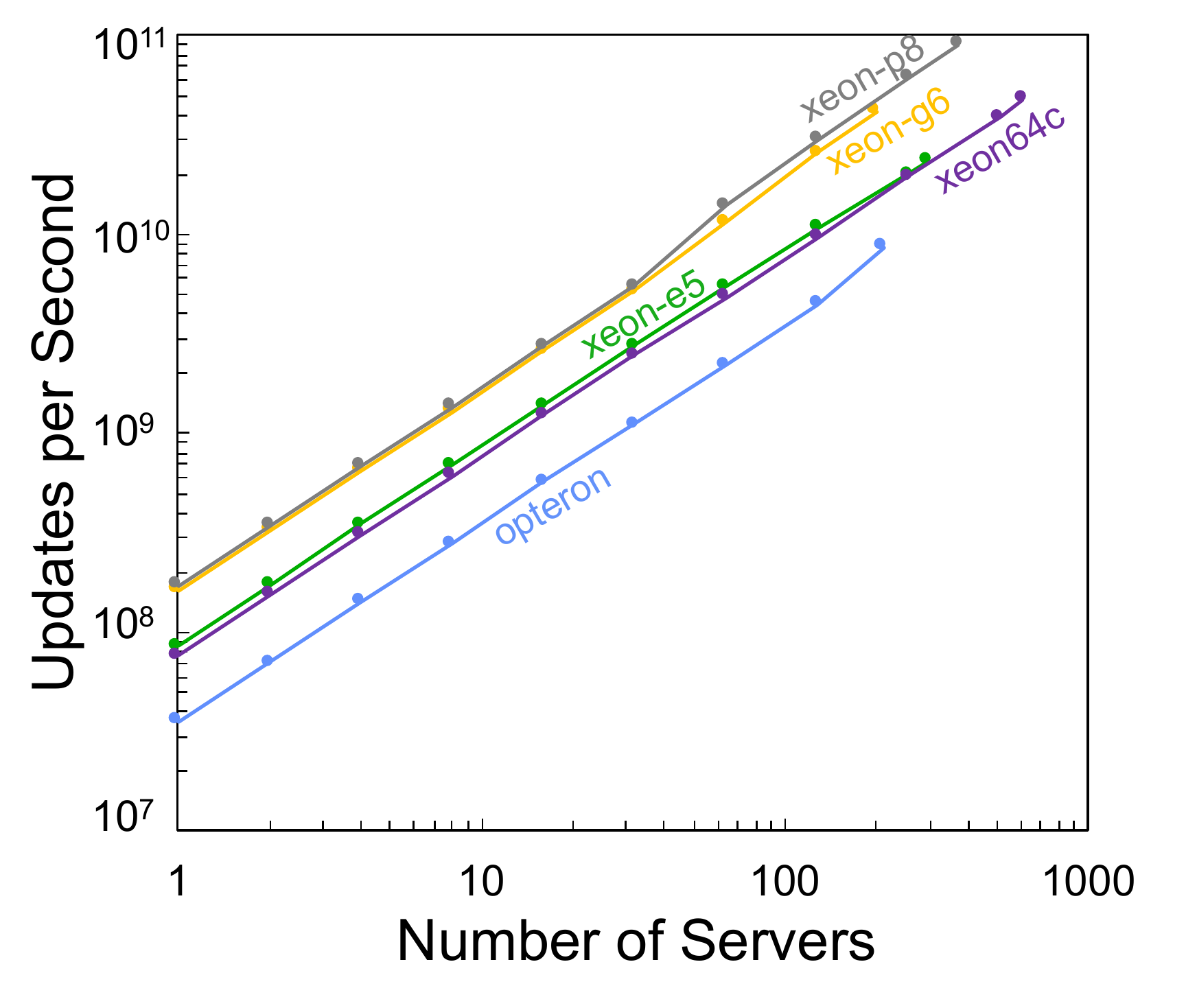}
\caption{Update rate as a function of number of servers for hierarchical GrapBLAS hypersparse matrices running on multiple nodes of different hardware types.}
\label{fig:UpdateRate-MultiNode}
\end{figure}

\begin{figure}[]
\centering
\includegraphics[width=\columnwidth]{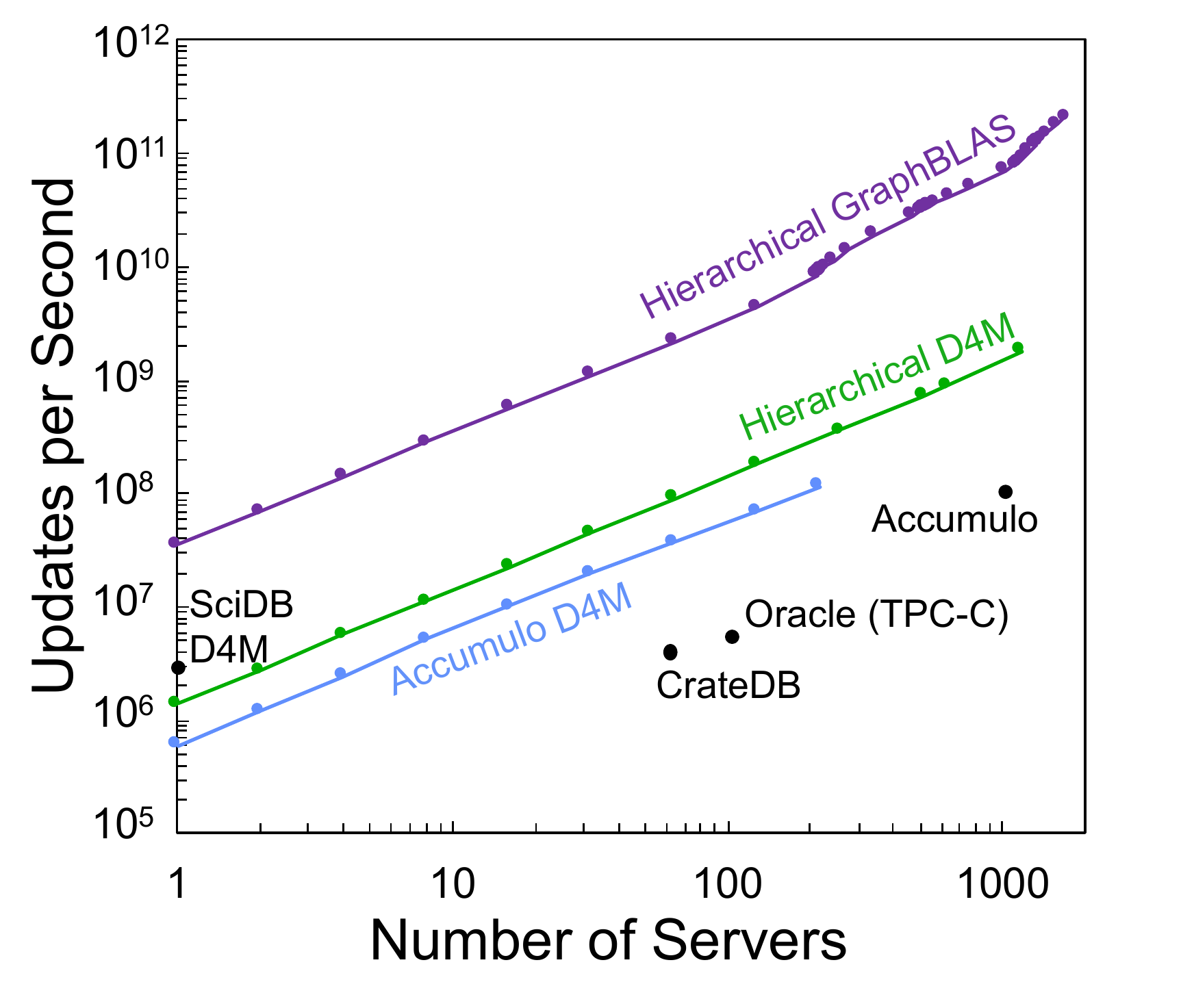}
\caption{Update rate as a function of number of servers for hierarchical GrapBLAS hypersparse matrices and other previous published work: Hierarchical D4M \cite{8547629}, Accumulo D4M \cite{kepner2014achieving}, SciDB D4M \cite{samsi2016benchmarking}, Accumulo \cite{sen2013benchmarking}, Oracle TPC-C benchmark, and CrateDB \cite{CrateDB}}
\label{fig:UpdateRate}
\end{figure}

Horizontal benchmarking examines the aggregate performance achieved by running on thousands of diverse compute nodes. The scalability of the hierarchical hypersparse matrices are tested using a power-law graph of 100,000,000 entries divided up into 1,000 sets of 100,000 entries.  These data were then simultaneously loaded and updated using a varying number of processes on varying number of nodes on the MIT SuperCloud.  The horizontal scaling for each type of hardware is shown in Figure~\ref{fig:UpdateRate-MultiNode}. Running on nearly two thousand  MIT SuperCloud nodes simultaneously achieved a sustained update rate of over 200,000,000,000 updates per second and is shown in Figure~\ref{fig:UpdateRate-MultiNode} with other technologies ranging such as in-memory index stores (D4M), NoSQL (Accumulo), NewSQL (SciDB, CrateDB), and SQL (Oracle) databases.   The achieved update rate of over 200,000,000,000 updates per second is significantly larger than the rate in prior published results \cite{kepner202075}. This capability allows the MIT SuperCloud to analyze extremely large streaming network data sets.

\section{Conclusion}

A variety of network, health, finance, and social applications can be enabled by hypersparse GraphBLAS matrices.  Hierarchical hypersparse GraphBLAS matrices enable rapid streaming updates while preserving algebraic analytic power and convenience.  The rate of these updates sets the bounds on performance in many contexts.  The GraphBLAS readily enable performance experiments on simultaneous diverse hardware via high-level language bindings to Matlab/Octave, Python, and Julia.  A peak single process performance was measured at over 4,000,000 updates per second.  The peak single node performance measured was 170,000,000 updates per second. The diverse MIT SuperCloud hardware used spans nearly a decade shows a 2x increase in single-core performance, a 3x increase in single process performance, and a 5x increase in single node performance over this era.   Running on nearly two thousand  MIT SuperCloud nodes simultaneously achieved a sustained update rate of over 200,000,000,000 updates per second.  Hierarchical hypersparse GraphBLAS allows the MIT SuperCloud to analyze extremely large streaming network data sets.  

\section*{Acknowledgement}

The authors wish to acknowledge the GraphBLAS community and the following individuals for their contributions and support: Bob Bond, Alan Edelman, Nathan Frey, Jeff Gottschalk, Tucker Hamilton, Chris Hill, Hayden Jananthan, Mike Kanaan, Tim Kraska, Charles Leiserson, Dave Martinez, Mimi McClure, Joseph McDonald, Christian Prothmann, John Radovan, Steve Rejto, Daniela Rus, Allan Vanterpool, Matthew Weiss, Marc Zissman.


%
%




\bibliographystyle{ieeetr}

\bibliography{HierGrBingest}
%

\end{document}